\documentclass[prb,twocolumn,showpacs,amsmath,amssymb,superscriptaddress]{revtex4}
\newcommand\beq{\begin{equation}}
\newcommand\eeq{\end{equation}}

\usepackage{dcolumn}
\usepackage{bm}

\usepackage{subfigure}
\usepackage{amsmath}
\usepackage{amsfonts}   
\usepackage{amssymb}
\usepackage[final]{graphicx}

\begin{document}


\title{Nearly-Perfect Non-Magnetic Invisibility Cloaking: Analytic Solutions and Parametric Studies}

\author{Giuseppe Castaldi}
\affiliation{Waves Group, Department of Engineering, University of Sannio, I-82100 Benevento, Italy}
\author{Ilaria Gallina}
\affiliation{Waves Group, Department of Engineering, University of Sannio, I-82100 Benevento, Italy}
\affiliation{Department of Environmental Engineering and Physics, University of Basilicata, I-85100 Potenza, Italy}
\author{Vincenzo Galdi}\email{vgaldi@unisannio.it}
\affiliation{Waves Group, Department of Engineering, University of Sannio, I-82100 Benevento, Italy}

\date{\today}

\begin{abstract}
Coordinate-transformation approaches to invisibility cloaking rely on the design of an anisotropic, spatially inhomogeneous ``transformation medium'' capable of suitably re-routing the energy flux around the region to conceal without causing any scattering in the exterior region. It is well known that the inherently magnetic properties of such medium limit the high-frequency scaling of practical ``metamaterial'' implementations based on subwavelength inclusions (e.g., split-ring resonators). Thus, for the optical range, {\em non-magnetic} implementations, based on approximate reductions of the constitutive parameters, have been proposed.

In this paper, we present an alternative approach to non-magnetic coordinate-transformation cloaking, based on the mapping from a {\em nearly-transparent}, anisotropic and spatially inhomogeneous virtual domain. We show that, unlike its counterparts in the literature, our approach is amenable to {\em exact analytic} treatment, and that its overall performance is comparable to that of a non-ideal (lossy, dispersive, parameter-truncated) implementation of standard (magnetic) cloaking.
\end{abstract}

\maketitle

\section{Introduction and Background}
\label{intro}

Invisibility of objects to an interrogating (electromagnetic, acoustic, elastic, quantum) wave illumination is a fascinating research topic of long-standing interest, with a wealth of intriguing theoretical and application-oriented implications. For instance, in electromagnetics (EM) engineering, ``invisible'' sources, scatterers and antennas have been investigated for several decades (see, e.g., Refs. \onlinecite{Kahn,Kerker,Chew,Alexopoulos,Kildal,Hoenders} for a sparse sampling).
However, during the last few years, interest in this topic has gained renewed momentum, under the suggestive association with the ``cloaking'' \cite{Schurig2} concept, mainly motivated by the rapid advances in the engineering of ``metamaterials'' with precisely controllable constitutive (e.g., anisotropy, spatial inhomogeneity, dispersion) properties. Among the most prominent approaches to (passive) invisibility cloaking, it is worth recalling those based on scattering cancellation \cite{Alu1,Silveirinha}, coordinate transformations \cite{Pendry,Leonhardt1,Leonhardt2,Schurig1} (experimentally demonstrated at microwave \cite{Schurig2} and visible \cite{Smolyaninov} frequencies), anomalous localized resonances \cite{Milton}, inverse design of scattering optical elements \cite{Hakansson}, and transmission-line networks \cite{Alitalo}. The reader is referred to Ref. \onlinecite{Alu2} (and references therein) for a recent comparative review of these various approaches.

In particular, the coordinate-transformation (also referred to as ``transformation-optics'' or ``transformation EM'') approach \cite{Schurig1,Pendry,Leonhardt1,Leonhardt2}, directly related to our investigation, relies on the formal invariance of Maxwell's equations under coordinate transformations, which allows a preliminary design in an auxiliary curved-coordinate space containing a ``hole,'' and its subsequent translation into a conventionally flat, Cartesian space filled by an anisotropic and spatially inhomogeneous ``transformation medium''
that suitably bends the ray trajectories so as to re-route the energy flux around the concealment region. The reader is referred to Refs. \onlinecite{Leonhardt3} and \onlinecite{Greenleaf2} (and references therein) for a recent collection and review of applications and theoretical aspects. Among the most interesting new twists and extensions of the basic cloaking idea above, it is worth mentioning the concepts of EM ``wormhole'' \cite{Greenleaf}, anti-cloak \cite{Chen3,Castaldi}, carpet cloak \cite{Li,Liu,Gabrielli,Valentine}, cloak at a distance \cite{Lai}, and open cloak \cite{Ma}, as well as the extensions to acoustic \cite{Cummer1,Pendry2}, elastic \cite{Milton2}, and quantum \cite{Zhang1} waves.

In view of the considerable complexity of the arising transformation media, EM modeling of such structures typically relies on heavily numerical (finite-element) simulations \cite{Cummer2}. Nevertheless, at least for canonical (e.g., cylindrical \cite{Kong2,Ruan} and spherical \cite{Kong1}) geometries, analytic full-wave approaches, based on suitable mappings of standard Fourier-Bessel or Mie expansions, have been developed. Remarkably, via these approaches, it was possible to prove analytically that in the ideal case (implying a lossless, anisotropic, spatially inhomogeneous transformation medium, with extreme values of the constitute parameters ranging from zero to infinity) the cloaking would be {\em perfect}, i.e., without any transmission into the concealment region and any external scattering, and may be, in principle, attained at {\em any} frequency -- not necessarily in the asymptotic high-frequency regime that the intuitive ray-bending picture would suggest. However, the inherent limitations arising from the unavoidable losses \cite{Kong2,Kong1}, dispersion \cite{Chen2,Yao}, perturbations \cite{Ruan} and simplifications/reductions \cite{Yan} in the constitutive parameters were also pointed out.

The first experimental verification of coordinate-transformation cloaking was achieved at microwave (X-band) frequencies \cite{Schurig2}, where the involved scales allowed the metamaterial fabrication via low-loss metallic split-ring-resonator inclusions. High-frequency scaling of this technological solution seems to be within reach for the low-THz region \cite{Tao}, but is limited by saturation effects \cite{Zhou}. Thus, for the visible range, an alternative route has been followed, based on the use of {\em non-magnetic} materials \cite{Cai1,Cai2,Cai3}, which has led to the experimental demonstration of an optical cloak \cite{Smolyaninov}. Non-magnetic approaches to coordinate-transformation cloaking are based on approximate reductions of the constitutive parameters that preserve the ray trajectories inside the cloak shell, at the expense of destroying the perfect impedance matching with the background medium, which can only be partially restored by suitable tweaking of the extra parameters available in quadratic \cite{Cai1,Cai2,Cai3} or higher-order \cite{Gallina,Zhang} coordinate transformations. The above parameter reductions also prevent application of the exact analytic approaches in Refs. \onlinecite{Kong2,Ruan,Kong1}, and thus their analytic modeling is limited to asymptotic (semiclassical) approximations \cite{Jacob}. 

In this paper, we propose a different approach to non-magnetic coordinate-transformation cloaking which, acknowledging the inherent limitations of realistic metamaterial implementations, instead of applying the coordinate transformation to a {\em perfectly transparent} virtual domain (as in standard transformation optics), considers a {\em weakly non-transparent} anisotropic and spatially inhomogeneous domain. With reference to a two-dimensional (cylindrical) scenario, we show that, via a judicious choice of the constitutive parameters and the coordinate transformation, it is possible to achieve a non-magnetic transformation-medium without resorting to approximate reductions, thereby maintaining the applicability of the computationally-effective and insight-providing exact analytic modeling (via generalization of the results in Refs. \onlinecite{Kong2} and \onlinecite{Ruan}). Moreover, via suitable parameter optimization at a given frequency, the inherently non-zero scattering can be minimized so that, when the unavoidable non-idealities (losses, dispersion, parameter truncations) of metamaterial implementations are taken into account, the overall performance becomes comparable to that of a standard (magnetic) cloak within broad parametric ranges.

Accordingly, the rest of the paper is laid out as follows. In Sec. \ref{Sec:Formulation}, we outline the problem formulation and the proposed strategy. In Sec. \ref{Sec:Analytic}, we derive the analytic solutions. In Sec. \ref{Sec:Results}, we present some representative results from our parametric studies, and compare them with those achievable via standard (magnetic) cloaking. Finally, in Sec. \ref{Sec:Conclusions}, we provide some concluding remarks and hints for future research.

%
\begin{figure}
\begin{center}
\includegraphics[width=8cm]{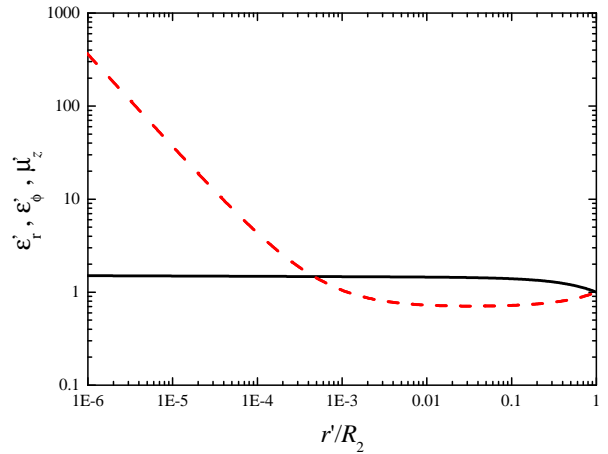}
\end{center}
\caption{(Color online) Virtual-domain-medium constitutive parameters $\varepsilon_r'=\varepsilon_\phi'$ (black solid) and $\mu_z'$ (red dashed) in (\ref{eq:start_eps_mu}), optimized for near-transparency at $R_2=3\lambda_0$ ($\gamma=3.41\cdot 10^{-3}$, $p=5.41\cdot 10^{-4}$, $\alpha=0.361$, see Sec. \ref{Sec:Optim}).}
\label{Figure1}
\end{figure}

\section{Problem Formulation and Proposed Strategy}
\label{Sec:Formulation}

\subsection{Virtual Space: Nearly-Transparent Domain}
In the two-dimensional (2-D) scenario of interest, we start considering an auxiliary virtual space $(x',y',z)$ featuring a circular cylindrical domain (infinite along $z$) of radius $R_2$, made of an anisotropic and radially inhomogeneous medium immersed in vacuum. We choose for the relative permittivity and permeability tensors ($\underline{\underline {\varepsilon}}'$ and $\underline{\underline {\mu}}'$, respectively) a rather general parametric form featuring several degrees of freedom and yet amenable to analytic solution of the relevant Helmholtz equation;
for the transverse-magnetic (TM) polarization (magnetic field parallel to the cylinder) of interest here, their relevant components (in the associated cylindrical coordinates $r', \phi, z$) are given by
\begin{subequations}
\begin{eqnarray}
\!\!\varepsilon_{r}'(r')\!\!\!&=&\!\!\varepsilon_{\phi}'(r')\!=\!\left(\frac{r'}{R_2}\right)^{-\gamma} \exp\left[\!-\alpha\left(\frac{r'}{R_2}-1\right)\!\right], \\
\mu_{z}'(r')\!\!\!&=&\!\!P(r') \left(\frac{r'}{R_2}\right)^{\gamma} \exp\left[\alpha\left(\frac{r'}{R_2}-1\right)\right]
\label{eq:start_eps_mu_11zz},
\end{eqnarray}
\label{eq:start_eps_mu}
\end{subequations}
where $\gamma>0$, $\alpha>0$, and
\beq
P(r')=1-p + p \frac{R_2}{r'},~~0\le p\le 1.
\label{eq:P}
\eeq
Note that the constitutive parameters in (\ref{eq:start_eps_mu}) are always positive, and locally matched with vacuum at the interface $r'=R_2$, whereas they may exhibit singular behaviors at $r'=0$,
\beq
\lim_{r'\rightarrow 0}\varepsilon_{r,\phi}'(r')=\infty,~~\lim_{r'\rightarrow 0}\mu_{z}'(r')=
\left\{
\begin{array}{lll}
\infty,~~0<\gamma<1,\\
\exp(-\alpha)p,~~\gamma=1,\\
0,~~\gamma>1.
\label{eq:sing1}
\end{array}
\right.
\eeq
%
\begin{figure*}
\begin{center}
\includegraphics[width=15cm]{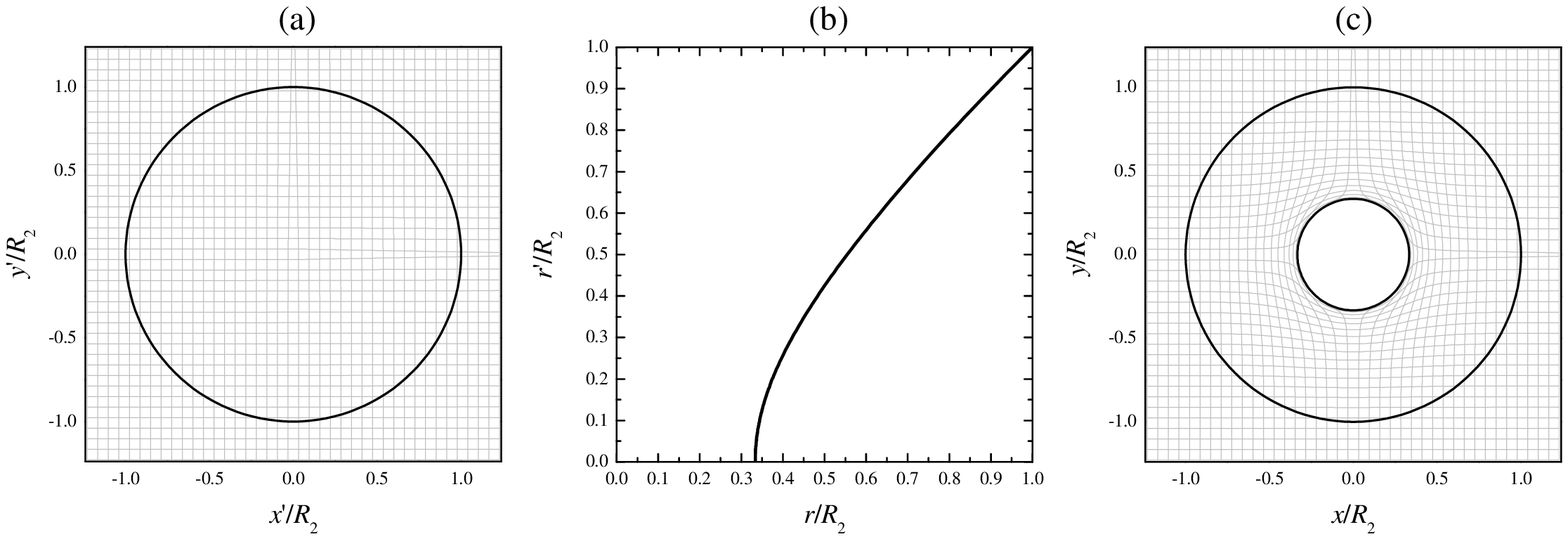}
\end{center}
\caption{(a) Ray trajectories (thin grey curves, traced along the positive $x'$ and $y'$ directions) in the virtual domain, with parameters as in Fig. \ref{Figure1}. (b) Coordinate transformation $r=g(r')$ vs. $r'=f(r)$ [cf. (\ref{functrancloak})--(\ref{functrancloakcontinuity})] for $R_1=R_2/3$. (c) Ray trajectories in the transformed domain. The black thick circles delimit the $r=R_1$ and $r=R_2$ cylindrical domains.}
\label{Figure2}
\end{figure*}
We highlight that, unlike the typical transformation-optics framework, our virtual domain is {\em not} perfectly transparent. Nevertheless, the constitutive relationships in (\ref{eq:start_eps_mu}) contain three adjustable parameters ($\alpha, \gamma, p$) that can be optimized (see Sec. \ref{Sec:Optim} below) for achieving a {\em nearly-transparent} response at a given frequency. Figure \ref{Figure1} shows  an example of constitutive parameters optimized for an electrical size $R_2=3\lambda_0$ (with $\lambda_0$ denoting the vacuum wavelength). The log-log scale utilized highlights the algebraic singular behavior of the permeability for $r'\rightarrow 0$, but is still not able to capture the extremely slow (in view of the chosen parameters) divergence of the permittivities.
For the same configuration, Fig. \ref{Figure2}(a) illustrates the ray tracing, obtained as in Ref. \onlinecite{Schurig1}. As a first indication of the nearly-transparent behavior of this optimized parametric configuration, one can observe the {\em practically straight} ray trajectories, with only a slight bending for those passing nearby the cylinder center.

\subsection{Coordinate Transformation}
\label{Sec:Transf}
Following the transformation-optics approach \cite{Schurig1,Pendry,Leonhardt1,Leonhardt2}, in order to achieve invisibility cloaking, we construct a cylindrical coordinate transformation, from the virtual space ($x',y',z$) to the actual physical space ($x,y,z$) of interest, 
\beq
r=g(r'),~~r'\le R_2,
\label{eq:mapping}
\eeq
which compresses the cylindrical nearly-transparent region $r'\le R_2$ into a concentric annulus $R_1\le r\le R_2$, i.e., satisfies the boundary conditions
\beq
g(0)=R_1,~~g(R_2)=R_2.
\label{eq:cloak_c}
\eeq
As previously mentioned, the non-flat metric underlying the transformation in (\ref{eq:mapping}) and (\ref{eq:cloak_c}) creates a ``hole'' of radius $R_1$, i.e., a region of space {\em effectively impenetrable} by the EM fields where an object can be concealed. Invoking the formal invariance of Maxwell's equations under coordinate transformations, the above behavior can be equivalently obtained in a {\em globally flat} space by filling up the transformed region $R_1\le r\le R_2$ with an anisotropic, spatially inhomogeneous ``transformation medium,'' whose relative permittivity and permeability tensors ($\underline{\underline {\varepsilon}}$ and $\underline{\underline {\mu}}$, respectively) are given by \cite{Schurig1,Pendry,Leonhardt1,Leonhardt2}
\beq
\left\{
\begin{array}{cc}
\underline{\underline {\varepsilon}}\\
\underline{\underline {\mu}}
\end{array}
\right\}
=\underline{\underline J} \cdot 
\left\{
\begin{array}{cc}
\underline{\underline {\varepsilon}}'\\
\underline{\underline {\mu}}'
\end{array}
\right\}
\cdot
\underline{\underline J}^T \left[\det\left(\underline{\underline J}\right)\right]^{-1},
\eeq
where $\underline{\underline J}=\partial{(x,y,z)}/\partial{(x',y',z)}$ is the Jacobian matrix of the transformation in (\ref{eq:mapping}), the superfix $^T$ indicates transposition, and the symbol $\det(\cdot)$ denotes the determinant.
For the TM polarization of interest here, the relevant (cylindrical) components can be directly found from (\ref{eq:start_eps_mu}) as \cite{Cai1}
\begin{subequations}
\begin{eqnarray}
\varepsilon_r(r)&=&\varepsilon_{r}'(r'){\dot g(r')}\frac{r'}{r},
\\
\varepsilon_\phi(r)&=&\frac{\varepsilon_{\phi}'(r')}{{\dot g(r')}} \frac{r}{r'},\\
\mu_z(r)&=&\frac{\mu_{z}'(r')}{{\dot g(r')}}\frac{r'}{r},
\label{eq:end_mu}
\end{eqnarray}
\label{eq:end_eps_mu}
\end{subequations}
where the overdot $\dot~$ denotes differentiation with respect to the argument, and $r'=f(r)$ [with $f(r)\equiv g^{-1}(r')$ denoting the inverse mapping]. In standard approaches to {\em non-magnetic} cloaking \cite{Cai1,Cai2,Cai3,Gallina,Zhang}, based on perfectly-transparent virtual domains (i.e., $\varepsilon_{r}'=\varepsilon_{\phi}'=\mu_{z}'=1$), reliance is made on the dependence of the ray trajectories 
on the products $\varepsilon_r \mu_z$ and $\varepsilon_\phi \mu_z$ in order to find a set of reduced parameters featuring $\mu_z=1$. As anticipated, this destroys the perfect transparency of the cloak, and prevents application of the exact analytic framework in Refs. \onlinecite{Kong2} and \onlinecite{Ruan}. In our approach, instead, we capitalize on the extra degrees of freedom available in order to enforce $\mu_z(r)=1$ in (\ref{eq:end_mu}), which yields the following first-order differential equation  
\beq
{\dot g(r')}-
\frac{r' \mu_{z}'(r')}{g(r')}
=
0.
\label{muzzcloakunity}
\eeq
A similar approach was used in Ref. \onlinecite{Luo} in order to obtain a spatially homogeneous axial permeability (or permittivity, for the transverse-electric polarization). However, relying on a vacuum (i.e., $\varepsilon'=\mu'=1$) virtual domain, such approach did not provide enough degrees of freedom. As a consequence, once the cloaking boundary conditions in (\ref{eq:cloak_c}) were enforced, the value of the transformed axial permeability was inevitably dictated by the shape factor $R_1/R_2$, and (most important) {\em always larger} than the vacuum value (i.e., $\mu_z>1$, for the case of interest here). 
In our approach, this limitation is overcome via the extra degrees of freedom endowed by the virtual-domain permeability $\mu_{z}'(r')$ [cf. Eqs. (\ref{eq:start_eps_mu_11zz}) and (\ref{eq:P})]. In particular, thanks to the functional form in (\ref{eq:start_eps_mu_11zz}), the differential equation in (\ref{muzzcloakunity}) admits a closed-form analytic solution as
\beq
g(r')
=
\sqrt{
R_1^2
+
R_2^2 h\!\left(\frac{r'}{R_2};\alpha,\gamma,p\right)
},
\label{functrancloak}
\eeq
where
\begin{eqnarray}
\!\!\!\!\!\!\!\!\!\!h&&\!\!\!\!\!\!\!\!\!\!\!\left(\frac{r'}{R_2};\alpha,\gamma,p\right)=
2\exp(-\alpha)\nonumber\\
&\times&\!\!\!\!
\left[
\left(\frac{1-p}{\gamma+2}\right)
\left(\frac{r'}{R_2}\right)^{\gamma+2}
M\!\!\left(
\gamma+2,\gamma+3,\frac{\alpha r'}{R_2}
\right)
\right.
\nonumber\\
\!\!\!&+&\!\!\!\!
\left.
\left(\frac{p}{\gamma+1}\right)
\left(\frac{r'}{R_2}\right)^{\gamma+1}
M\!\!\left(
\gamma+1,\gamma+2,\frac{\alpha r'}{R_2}
\right)
\!\right]\!,
\label{eq:h}
\end{eqnarray}
with $M(\cdot,\cdot,\cdot)$ denoting the confluent hypergeometric function [Eq. (13.1.2) in Ref. \onlinecite{Abramowitz}]. Note that, in light of the assumed constraints ($\alpha>0$, $\gamma>0$, $0\le p\le 1$), the function $h(r'/R_2;\alpha,\gamma,p)$ in (\ref{eq:h}) is always positive, which ensures that the mapping $g(r')$ in (\ref{functrancloak}) is always real. Moreover, in (\ref{functrancloak}) and (\ref{eq:h}), the arising integration constant has been exploited to enforce the first boundary condition in (\ref{eq:cloak_c}); the second boundary condition, instead, yields
\beq
h\left(1;\alpha,\gamma,p\right)
=
1-\left(\frac{R_1}{R_2}\right)^2,
\label{functrancloakcontinuity}
\eeq
which can be satisfied by properly tweaking the parameters $\alpha$, $\gamma$, and $p$. In particular, in view of the {\em linear} dependence involved, Eq. (\ref{functrancloakcontinuity}) may be straightforwardly solved with respect to $p$. However, this seemingly simplest approach is not necessarily the most effective in a broader perspective of achieving a nearly-transparent response. In fact, our parametric studies (see Sec. \ref{Sec:Optim} below) indicate that it is generally more convenient to satisfy the constraint in (\ref{functrancloakcontinuity}) by fixing $\alpha$ (via numerical solution), and exploit the parameters $\gamma$ and $p$ for minimizing the scattering response.

It should be noted that, since $\mu_{z}'(r')$ in (\ref{eq:start_eps_mu_11zz}) and $g(r')$ in (\ref{functrancloak}) are always positive, the constraint in (\ref{muzzcloakunity}) also implies $\dot g(r')>0$, and hence the invertibility of the coordinate mapping. However, the inverse mapping $f(r)=g^{-1}(r')$ cannot generally be calculated analytically. Nevertheless, as can be observed from the example in Fig. \ref{Figure2}(b), the mapping behavior is fairly regular, and therefore its numerical inversion does not pose any problem.
Figure \ref{Figure2}(c) shows the ray trajectories obtained by transforming [via the mapping in Fig. \ref{Figure2}(b)] those in the virtual space [cf. Fig. \ref{Figure2}(a)], from which the ray bending around the interior region ($r<R_1$) to conceal (typical of coordinate-transformation-based cloaking) is fairly evident.

\subsection{Real Space: Non-Magnetic Transformation Medium}
By substituting (\ref{eq:start_eps_mu}) and $r'=f(r)$ in (\ref{eq:end_eps_mu}), and taking into account (\ref{muzzcloakunity}), we readily obtain the explicit expressions of the constitutive parameters of the desired {\em non-magnetic} transformation medium in the real space, 
\begin{subequations}
\begin{eqnarray}
\!\!\!\!\!\!\!\!\!\!\varepsilon_{r}(r)\!\!\!&=&\!\!\!\frac{R_2^{\gamma} \left[f(r)\right]^{1-\gamma}}{r  {\dot f(r)}} 
\exp\left\{\!-\alpha\!\left[\!\frac{f(r)}{R_2}-1\!\right]\!\right\},
\label{eq:eps1}\\
\varepsilon_{\phi}(r)\!\!\!&=&\!\!\!\frac{R_2^{\gamma} r {\dot f(r)}}{\left[f(r)\right]^{\gamma+1}} 
\exp\left\{-\alpha\left[\frac{f(r)}{R_2}-1\right]\right\},
\label{eq:eps2}\\
\mu_{z}(r)\!\!\!&=&\!\!\! 1.
\label{eq:end_eps_mu_11zzrho}
\end{eqnarray}
\label{eq:end_eps_murho}
\end{subequations}
Note that, since $f(r)$ and $\dot f(r)$ are always positive in the transformed region $R_1\le r\le R_2$, both permittivity components in (\ref{eq:eps1}) and (\ref{eq:eps2}) are likewise positive, thereby yielding a {\em double positive} medium. Moreover, like the virtual-space medium in (\ref{eq:start_eps_mu}), our non-magnetic transformation medium in (\ref{eq:end_eps_murho}) is locally matched with vacuum at $r=R_2$. 
In view of the singular behavior exhibited by the virtual-domain medium at $r'=0$ [cf. (\ref{eq:sing1})], it is interesting to investigate the behavior of the transformation medium at the image point $r=R_1$.
Recalling that $M(a,b,0)=1$ [cf. Eq. (13.1.2) in Ref. \onlinecite{Abramowitz}], we obtain from (\ref{eq:h}), in the limit $r'\rightarrow 0$,
\beq
h\left(\frac{r'}{R_2};\alpha,\gamma,p\right)\sim
2\exp(-\alpha)
\left(
\frac{\chi}{\gamma-s+2}\right)
\left(\frac{r'}{R_2}\right)^{\gamma-s+2},
\label{eq:h_app}
\eeq
where
\beq
\chi=\left\{
\begin{array}{ll}
1, ~ p=0,\\ 
p,~ p\ne0,
\end{array}
\right.
~~s=
\left\{
\begin{array}{ll}
0, ~ p=0,\\ 
1,~ p\ne0. 
\end{array}
\right.
\eeq
By substituting (\ref{eq:h_app}) into (\ref{functrancloak}), and approximating the square root via its first-order Taylor expansion, we then obtain
\begin{subequations}
\begin{eqnarray}
g(r')&\sim& R_1\sqrt{
1 + \frac{2 \exp(-\alpha) \chi R_2^{s-\gamma}  (r')^{\gamma-s+2}}{(\gamma-s+2)R_1^2}}\nonumber\\
&\sim& 
R_1 + 
\frac{\exp(-\alpha) \chi R_2^{s-\gamma}  (r')^{\gamma-s+2}}{(\gamma-s+2)R_1},
\label{eq:g_app}
\end{eqnarray}
and hence, via differentiation,
\beq
{\dot g(r')}
\sim 
 \exp(-\alpha)\chi
R_1^{-1} 
R_2^{s-\gamma} (r')^{\gamma-s+1}.
\label{eq:gp}
\eeq
\end{subequations}
Recalling that $\dot f(r)=1/\dot g(r')$ and $r'=f(r)$, and substituting into (\ref{eq:eps1}) and (\ref{eq:eps2}), we then obtain 
\begin{subequations}
\begin{eqnarray}
\varepsilon_{r}(r)& \sim &
\frac{  
\chi
R_2^{s}  
}{R_1^2}
\left[f(r)\right]^{2-s},\\
\varepsilon_{\phi} (r)& \sim &  
\frac{
\exp(2\alpha)
R_1^2
R_2^{2\gamma-s}
}
{
\chi
} \left[f(r)\right]^{s-2\gamma-2},
\label{eq:end_eps_mu_11zzrho_zerolimit}
\end{eqnarray}
\label{eq:end_eps_murho_zerolimit}
\end{subequations}
from which, recalling that $0\le s\le 1$ and $\gamma>0$, it finally follows
\beq
\varepsilon_{r}(R_1)=0,~~
\lim_{r\rightarrow R_1} \varepsilon_{\phi}(r)=\infty.
\label{eq:sing_eps}
\eeq
%
\begin{figure}
\begin{center}
\includegraphics[width=8cm]{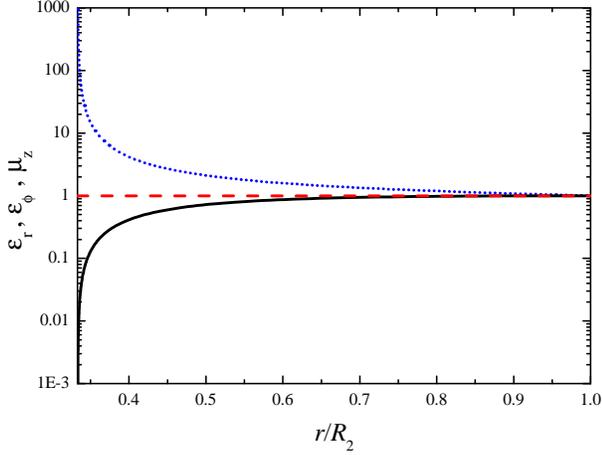}
\end{center}
\caption{(Color online) Non-magnetic transformation-medium constitutive parameters $\varepsilon_r$ (black solid), $\varepsilon_\phi$ (blue dotted) and $\mu_z$ (red dashed) in (\ref{eq:end_eps_murho}), obtained from the nearly-transparent virtual domain in Fig. \ref{Figure1} via the mapping in Fig. \ref{Figure2}(b).}
\label{Figure3}
\end{figure}
Thus, as in standard (magnetic) cloaking, at the inner interface $r=R_1$, the permittivities exhibit either zero or infinite values. 
Figure \ref{Figure3} shows the transformation-medium constitutive parameters obtained from the (optimized) virtual-domain medium in Fig. \ref{Figure1}, via the coordinate mapping in Fig. \ref{Figure2}(b), and corresponding to the ray trajectories in Fig. \ref{Figure2}(c).

It it worth pointing out that, in principle, different choices of the parameters and functional forms of the virtual-domain medium (\ref{eq:start_eps_mu}) are possible, yielding {\em finite} variation ranges of the resulting transformation media. Our choice here was only aimed at facilitating direct comparison with standard (magnetic) cloaks. Nevertheless, the effects of unavoidable parameter truncations are investigated in our parametric studies (see Sec. \ref{Sec:RealSpace1} below).

\section{Analytic Solutions}
\label{Sec:Analytic}

The EM response of the cloaking configuration of interest can be computed analytically by generalizing the Fourier-Bessel-type approach in Refs. \onlinecite{Kong2} and \onlinecite{Ruan}. In what follows, we address such generalization for the case of TM-polarized plane-wave illumination with unit-amplitude $z-$directed magnetic field and time-harmonic [$\exp(-i\omega t)$] dependence, by suitable coordinate mapping [via the inverse of (\ref{eq:mapping})] of the virtual-space solution.

\subsection{Virtual Space}

It is expedient to expand the incident plane wave $H_z^i$ (assumed to impinge from the positive $x'$ direction) and the scattered field $H_z^s$ into  Fourier-Bessel series,
\begin{subequations}
\begin{eqnarray}
H^i_z(r',\phi)&=&\exp(ik_0x')\nonumber\\
&=&\sum_{m=-\infty}^{\infty}i^m J_m\left(k_0r'\right)\exp(im\phi),
\label{eq:PW}\\
H^s_z(r',\phi)&=&\sum_{m=-\infty}^{\infty}c_m H_m^{(1)}\left(k_0r'\right)\exp(im\phi),
\end{eqnarray}
\label{eq:H_is}
\end{subequations}
where $k_0=\omega/c=2\pi/\lambda_0$ denotes the vacuum wavenumber (with $c$ denoting the speed of light in vacuum), $J_m(\cdot)$ and $H_m^{(1)}(\cdot)$ are the $m$-th order Bessel and Hankel functions of first kind, respectively (cf. Sec. 9.1 in Ref. \onlinecite{Abramowitz}), and $c_m$ are unknown coefficients. The magnetic field $H^t_z$ transmitted inside the cylindrical domain $r'<R_2$, ruled by the Helmholtz equation \cite{Kong2}
\begin{eqnarray}
\!\!\!\!\!\!\!\!\!\frac{1}{r' \mu_{z}'(r')} 
&&
\!\!\!\!\!\!\!\!\!\!\!\!\frac{\partial}
{\partial r'}
\left[
\frac{r'}{\varepsilon_{\phi}'(r')} \frac{\partial}{\partial r'}
\right]H^t_z(r',\phi)\nonumber\\
\!\!\!\!\!\!\!\!&+&
\!\!\!\!\left[
\frac{1}{r'^2\varepsilon_{r}'(r')\mu_{z}'(r')}
\frac{
\partial^2 
}{\partial \phi^2}
\!+\! k_0^2\right]\!\!
H^t_z(r',\phi)
=
0,
\label{eqMax}
\end{eqnarray}
can be likewise expanded into this type of series
\begin{eqnarray}
H^t_{z}(r',\phi)\!\!\!&=&\!\!\!\sum_{m=-\infty}^{\infty}\left[a_m \Psi_m^{(1)}(k_0r')+ b_m \Psi_m^{(2)}(k_0r')\right]\nonumber\\ 
&\times&\exp(i m \phi),~~r'\le R_2.
\label{eq:Ht}
\end{eqnarray}
In (\ref{eq:Ht}), $a_m$ and $b_m$ are unknown coefficients, while $\Psi_m^{(1,2)}(\cdot)$ denote independent solutions of the radial Helmholtz equations
\begin{eqnarray}
\left\{
\frac{
d^2
}
{d r'^2}
+
\left[
\frac{1}{r'}-\frac{1}{\varepsilon'_{\phi}(r')}\frac{d \varepsilon'_{\phi}(r')}{d r'} 
\right]
\frac{d}{d r'}\right\}
\Psi_m(k_0r')
\nonumber\\
+
\left[
k_0^2
\mu'_{z}(r')
\varepsilon'_\phi(r')
-
\frac{m^2 \varepsilon'_{\phi}(r')}{r'^2 \varepsilon'_{r}(r')}
\right]
\Psi_m(k_0r')
=
0.
\label{eq:Helmholtz1}
\end{eqnarray}
It can be shown (see Appendix \ref{Sec:AppA} for details) that, for the constitutive parameters in (\ref{eq:start_eps_mu}), these solutions can be expressed in closed form as
\begin{eqnarray}
\Psi_m^{(1,2)}(k_0r')&=&
\left(\frac{r'}{R_2}\right)^{\frac{\nu_m-\gamma}{2}} 
\exp\left[-\frac{\left(\alpha+\xi R_2\right)r'}{2R_2}\right]\nonumber\\
&\times&
\left\{
\begin{array}{ll}
M(\zeta_m,\nu_m+1,\xi r')\\
U(\zeta_m,\nu_m+1,\xi r')
\end{array}
\right\},
\label{eqSolTM1}
\end{eqnarray}
where (the already defined) $M(\cdot,\cdot,\cdot)$ and $U(\cdot,\cdot,\cdot)$ are confluent hypergeometric functions [cf. Eqs. (13.1.2) and (13.1.3), respectively, in Ref. \onlinecite{Abramowitz}], and
\begin{subequations}
\begin{eqnarray}
\!\!\!\!\!\!\!\!\!\!\!\nu_m\!\!&=&\!\!\sqrt{\gamma^2+4m^2},
\label{eq:num}\\
\!\!\!\!\!\!\!\!\!\!\!\xi\!\!&=&\!\!R_2^{-1}\sqrt{\alpha^2-4 (1-p) k_0^2 R_2^2},~~0\le\arg(\xi)<\pi,
\label{eq:xi}\\
\!\!\!\!\!\!\!\!\!\!\!\!\!\!\zeta_m
\!\!&=&\!\!
\frac{
 \xi(\nu_m+1)R_2
+
\alpha(\gamma+1)
-
2 p k_0^2 R_2^2 
}
{
2 \xi R_2
}.
\label{eq:zetam}
\end{eqnarray}
\end{subequations}
Recalling that the wavefunctions $\Psi_m^{(2)}$ exhibit a singular behavior at $r'=0$ (see Appendix \ref{Sec:AppB}), the field-finiteness condition yields $b_m=0$ in (\ref{eq:Ht}). The remaining unknown expansion coefficients ($a_m$ and $c_m$) can be computed by enforcing the continuity of the magnetic and electric field tangential components at the interface $r=R_2$, viz.,
\begin{subequations}
\begin{eqnarray}
H_z^t(R_2,\phi)&=&H_z^i(R_2,\phi)+H_z^s(R_2,\phi),
\label{eq:bc1}\\
E_{\phi}^t(R_2,\phi)&=&E_{\phi}^i(R_2,\phi)+E_{\phi}^s(R_2,\phi),
\label{eq:bc2}
\end{eqnarray}
\label{eq:bbc}
\end{subequations}
where the tangential electric fields readily follow from the curl Maxwell equation
\beq
E_{\phi}(r',\phi)=\frac{1}{i\omega\varepsilon_0\varepsilon'_{\phi}(r')}\frac{\partial H_z(r',\phi)}{\partial r'}.
\label{eq:Ep}
\eeq
By substituting the Fourier expansions (\ref{eq:H_is}) and (\ref{eq:Ht}) into (\ref{eq:bbc}) [with (\ref{eq:Ep})], we obtain a doubly countable infinity of linear equations,
\begin{subequations}
\begin{eqnarray}
a_m \Psi^{(1)}_m(k_0R_2)=i^m J_m(k_0 R_2) + c_m H_m^{(1)}(k_0 R_2),
\label{eq:Cloak_3bRho}
\\
a_m {\dot\Psi^{(1)}_m(k_0R_2)}=i^m {\dot J_m(k_0 R_2)} + c_m {\dot H_m^{(1)}(k_0 R_2)},
\label{eq:Cloak_4bRho}
\end{eqnarray}
\label{eq:CloakTotRho}
\end{subequations}
which can be solved in a straightforward fashion, yielding
\begin{subequations}
\begin{eqnarray}
a_m 
&=&
\frac{
2 i^{m+1}
}
{
\pi k_0R_2 W_{\left[H_m^{(1)},\Psi_m^{(1)}\right]}
},\\ 
c_m
&=&
\frac{
-i^m
W_{\left[J_m,\Psi_m^{(1)}\right]}
}
{
W_{\left[H_m^{(1)},\Psi_m^{(1)}\right]}
}, \label{eq:cm}
\end{eqnarray}
\label{eq:amcm}
\end{subequations}
where $W_{\left[F,G\right]}$ represents the Wronskian of the functions $F$ and $G$ evaluated at $k_0R_2$,
\beq
W_{\left[F,G\right]}=F\left(k_0R_2\right){\dot G\left(k_0R_2\right)}-{\dot F\left(k_0R_2\right)}G\left(k_0R_2\right).
\eeq

\subsection{Real Space}
\label{Sec:RealSpace}
We can now address the solution in the real space $(r,\phi,z)$, by generalizing the approach in Refs. \onlinecite{Kong2} and \onlinecite{Ruan}. First, we note that, since the coordinate transformation in (\ref{eq:mapping}) is restricted within the cloak shell $R_1\le r\le R_2$, the fields in the (vacuum) exterior region $r>R_2$ admit the same expressions as in (\ref{eq:H_is}) (with $r$ substituting $r'$). For the same reason, the field transmitted into the (vacuum) concealment region $r\le R_1$, can be expanded in terms of a standard Fourier-Bessel series,
\begin{subequations}
\beq
H^t_z(r,\phi)=\sum_{m=-\infty}^{\infty} d_m J_m(k_0 r) \exp(i m \phi),~~r\le R_1,
\eeq
where $d_m$ are unknown coefficients, and the field-finiteness condition has been enforced. Following Refs. \onlinecite{Kong2} and \onlinecite{Ruan}, the field transmitted into the cloak shell $R_1\le r \le R_2$ is obtained by (inverse) coordinate mapping [via (\ref{eq:mapping})] of the virtual-space solution (\ref{eq:Ht}), viz.,
\begin{eqnarray}
H^t_z(r,\phi)&=&
\sum_{m=-\infty}^{\infty}
\left[
a_m \psi_m^{(1)}(k_0r) 
+
b_m \psi_m^{(2)}(k_0r) 
\right]\nonumber\\
&\times&\exp(i m \phi),~~R_1\le r \le R_2, 
\end{eqnarray}
\label{eq:Ht2}
\end{subequations}
where $a_m$ and $b_m$ are unknown expansion coefficients, and
\beq
\psi_m^{(1,2)}(k_0r)=\Psi_m^{(1,2)}\left[k_0f(r)\right].
\label{eq:psi12}
\eeq
Once again, the four sets of unknown expansion coefficients ($a_m, b_m, c_m, d_m$) can be computed by enforcing the continuity of the tangential fields, this time at the interfaces $r=R_1$ and $r=R_2$. While the continuity at the outer interface ($r=R_2$) does not pose any particular problem, much more involved is dealing with the inner interface ($r=R_1$), in view of the singular behavior exhibited by the wavefunctions $\psi_m^{(2)}$ in (\ref{eq:psi12}) [cf. Eq. (13.1.3) in Ref. \onlinecite{Abramowitz}, and Appendix \ref{Sec:AppB}] and the azimuthal permittivity $\varepsilon_{\phi}$ in (\ref{eq:eps2}) [cf. (\ref{eq:sing_eps})]. In order to circumvent this problem, as in Refs. \onlinecite{Kong2} and \onlinecite{Ruan}, we therefore follow a limiting approach, slightly shifting the inner cloak boundary to $r=R_1+\Delta$, where $\Delta$ denotes a small quantity (which eventually we let tend to zero). Accordingly, we obtain
\begin{widetext}
\begin{subequations}
\begin{eqnarray}
d_m J_m[k_0 (R_1+\Delta)]&=&a_m \psi_m^{(1)}[k_0(R_1+\Delta)]+b_m \psi_m^{(2)}[k_0(R_1+\Delta)],
\label{eq:Cloak_3a}
\\
d_m {\dot J_m[k_0 (R_1+\Delta)]}&=&\frac{a_m {\dot\psi_m^{(1)}}[k_0(R_1+\Delta)]+b_m {\dot\psi_m^{(2)}[k_0(R_1+\Delta)]}}{\varepsilon_{\phi}(R_1+\Delta)},
\label{eq:Cloak_4a}
\\
a_m \psi_m^{(1)}(k_0 R_2)+b_m \psi_m^{(2)}(k_0 R_2)&=&i^m J_m(k_0 R_2) + c_m H_m^{(1)}(k_0 R_2),
\label{eq:Cloak_3b}
\\
a_m {\dot\psi_m^{(1)}(k_0 R_2)}+b_m {\dot\psi_m^{(2)}(k_0 R_2)}&=&i^m {\dot J_m(k_0 R_2)} + c_m {\dot H_m^{(1)}(k_0R_2)}.
\label{eq:Cloak_4b}
\end{eqnarray}
\label{eq:CloakTot}
\end{subequations}
\end{widetext}
It can be shown (see Appendix \ref{Sec:AppB} for details) that, in the limit $\Delta\rightarrow 0$, Eqs. (\ref{eq:Cloak_3a}) and (\ref{eq:Cloak_4a}) yield
\begin{subequations}
\begin{eqnarray}
b_m
&\sim&
a_m
\Omega_1\left(\frac{\Delta}{R_2}\right)^{\frac{\nu_m}{\gamma-s+2}},
\label{eq:Sol_4aappenAA}\\
d_m
&\sim&
a_m
\Omega_2
\left(\frac{\Delta}{R_2}\right)^{\frac{\gamma+\nu_m}{2(\gamma-s+2)}},
\label{eq:sol_3aappenAA}
\end{eqnarray}
\label{eq:CloakTotappenSolAA}
\end{subequations}
where $\Omega_{1,2}$ are irrelevant constants.
Recalling that $0\le s\le 1$, $\gamma>0$ and $\nu_m>0$, it is readily realized that the expansion coefficients $b_m$ and $d_m$ in (\ref{eq:CloakTotappenSolAA}) vanish as $\Delta\rightarrow 0$. As a consequence, recalling (\ref{eq:psi12}) and that $f(R_2)=R_2$, Eqs. (\ref{eq:Cloak_3b}) and (\ref{eq:Cloak_4b}) are readily recognized to become identical to (\ref{eq:Cloak_3bRho}) and (\ref{eq:Cloak_4bRho}), respectively, and therefore the remaining unknown coefficients $a_m$ and $c_m$ are still given by (\ref{eq:amcm}).

To sum up, the above analytic solutions resemble those obtained for the standard (magnetic) cloak \cite{Kong2} in the {\em exact suppression} of the field transmitted into the concealment region (i.e., $d_m=0$). The expectable differences show up in the {\em non-zero} scattering coefficients $c_m$, which are directly inherited from the nearly-transparent medium (\ref{eq:start_eps_mu}) in the virtual space. 

In what follows, we show that, via judicious optimization at a given frequency, the scattering response can be effectively reduced so that the overall performance of the proposed nonmagnetic cloak becomes comparable to that of a standard (magnetic) cloak when the unavoidable non-idealities (parameter truncations, losses, dispersion) are taken into account.
%
\begin{figure}
\begin{center}
\includegraphics[width=8cm]{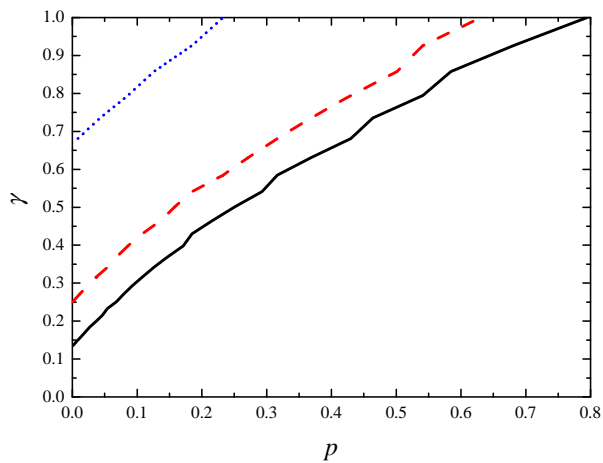}
\end{center}
\caption{(Color online) Curves in the ($\gamma,p$) plane yielding $\alpha=0$ roots of (\ref{functrancloakcontinuity}), for the parameters in Fig. \ref{Figure1} and shape-factor values $R_1/R_2=1/4, 1/3, 1/2$ (black-solid, red-dashed, blue-dotted, respectively). The regions below the curves are associated with positive $\alpha$ values, and thus represent the admissible $(\gamma,p)$ search spaces in the scattering-width minimization problem.}
\label{Figure4}
\end{figure}

\section{Representative Results}
\label{Sec:Results}

We now move on to the presentation of the salient results from an extensive series of parametric studies, starting with the virtual-domain parameter optimization, and continuing with the performance comparison between the proposed approach and standard (magnetic) cloaking.

\subsection{Virtual Space: Optimization for Near-Transparency}
\label{Sec:Optim}
As previously mentioned, the inherently non-zero scattering response of our proposed non-magnetic cloak is directly inherited by the non-transparent virtual domain in (\ref{eq:start_eps_mu}). In our study, such response is compactly parameterized in terms of the total scattering cross-sectional width per unit length \cite{Silveirinha}
\beq
Q_s=\frac{4}{k_0} \sum_{m=-\infty}^{\infty} |c_m|^2,
\label{eq:Qs}
\eeq
where the scattering coefficients $c_m$ are given by (\ref{eq:cm}). Our approach is based on the minimization of the scattering width in (\ref{eq:Qs}), at a given frequency, by acting on the free parameters ($\alpha, \gamma, p$) available in the virtual-domain-medium constitutive parameters (\ref{eq:start_eps_mu}). Clearly, different observables (e.g., more directly tied to the near-field or angular distributions) may be considered, giving rise to different optimized configurations. It is also worth recalling that the above parameters are actually constrained via the cloak condition in (\ref{functrancloakcontinuity}).
%
\begin{figure}
\begin{center}
\includegraphics[width=8cm]{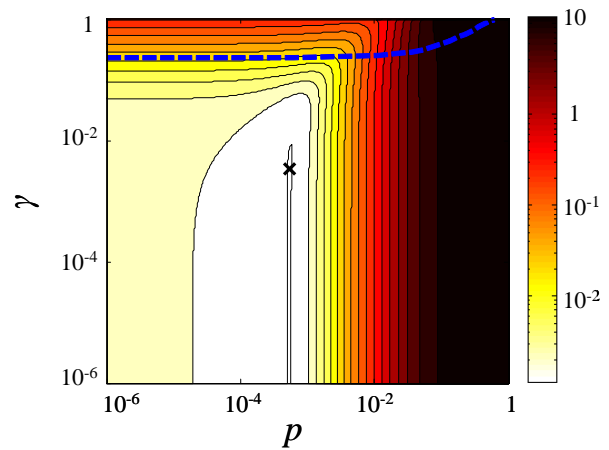}
\end{center}
\caption{(Color online) Contour plot of the scattering width in (\ref{eq:Qs}) as a function of $\gamma$ and $p$, for the parameters in Fig. \ref{Figure1} and shape factor $R_1/R_2=1/3$. The blue-dashed curve delimits the admissible search space (cf. Fig. \ref{Figure4}), whereas the cross indicates the minimum.}
\label{Figure5}
\end{figure}
In our numerical studies, we found it most effective to satisfy this constraint by fixing $\alpha$ via numerical solution of (\ref{functrancloakcontinuity}), and exploit the parameters $\gamma$ and $p$ for minimizing $Q_s$. Figure \ref{Figure4} shows, for representative values of the shape factor $R_1/R_2$, the $(\gamma,p)$ parametric ranges for which (\ref{functrancloakcontinuity}) admits a solution $\alpha>0$ (consistent with the model assumptions), which constitute the search spaces in our minimization problem \cite{footnote1}. 
In view of the reduced number of parameters involved and the computationally-effective analytic modeling, such minimization can be readily pursued via exhaustive parameter scanning. For a fixed frequency and shape factor, Fig. \ref{Figure5} shows the scattering width in (\ref{eq:Qs}) as a function of the two free parameters $\gamma$ and $p$, within the above defined admissible ranges, from which a rather broad minimum is identified. The corresponding parameter configuration yields the constitutive relationships shown in Fig. \ref{Figure1} and the associated (practically straight) ray trajectories in Fig. \ref{Figure2}(a). For the same ideal (lossless, non-truncated) parameter configuration, Fig. \ref{Figure6} shows the real part of the magnetic field map (for plane-wave excitation) computed via the expansions in (\ref{eq:H_is}) and (\ref{eq:Ht}) [with (\ref{eq:amcm})], from which the near-transparency is evident. 

%
\begin{figure}
\begin{center}
\includegraphics[width=8cm]{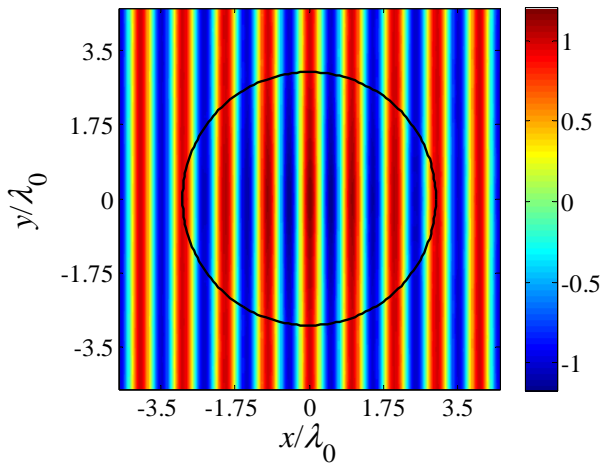}
\end{center}
\caption{(Color online) Magnetic field (real part) map in the virtual space, for unit-amplitude TM-polarized plane-wave illumination (impinging along the positive $x-$direction), and with ideal (lossless, non-truncated) parameters as in Fig. \ref{Figure1}.}
\label{Figure6}
\end{figure}

%
\begin{figure}
\begin{center}
\includegraphics[width=8cm]{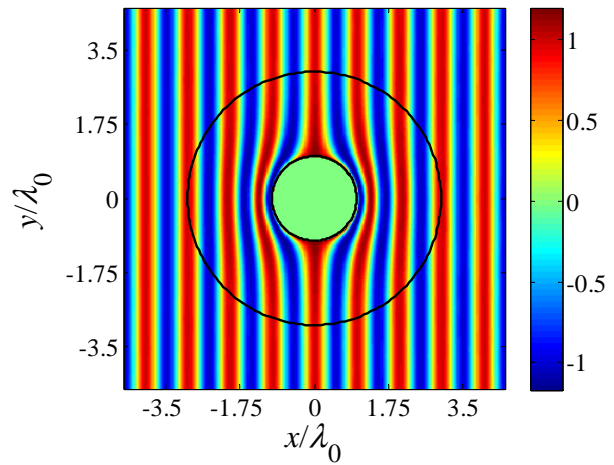}
\end{center}
\caption{(Color online) As in Fig. \ref{Figure6}, but for the real domain, with ideal (lossless, non-truncated) parameters as in Fig. \ref{Figure3}.}
\label{Figure7}
\end{figure}

\subsection{Real Space: Comparison with Standard (Magnetic) Cloak}
\label{Sec:RealSpace1}
Referring to the non-magnetic ideal (lossless, non-truncated) constitutive parameters in Fig. \ref{Figure3} and associated ray trajectories in Fig. \ref{Figure2}(c) [derived from the above optimized virtual-domain via the mapping in Fig. \ref{Figure2}(b)], Fig. \ref{Figure7} shows the corresponding field map, computed via the expansions in (\ref{eq:H_is}) and (\ref{eq:Ht2}) [with (\ref{eq:amcm}) and (\ref{eq:CloakTotappenSolAA})], which highlights the cloaking effect accompanied by a {\em very weak} scattering (identical to that in Fig. \ref{Figure6}).
%
\begin{figure}
\begin{center}
\includegraphics[width=8cm]{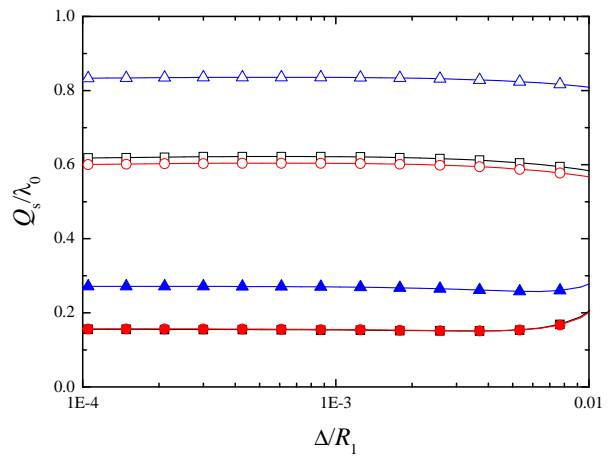}
\end{center}
\caption{(Color online) Parameters as in Fig. \ref{Figure3}, but truncated at $r=R_1+\Delta$. Total scattering width (normalized with respect to the vacuum wavelength), as a function of the truncation parameter $\Delta/R_1$, for loss-tangent values $\tan\delta=0, 10^{-3}, 10^{-2}$ (full black squares, red circles, and blue triangles, respectively). As a reference, the corresponding values attainable from a standard (magnetic) cloak [cf. (\ref{eq:end_eps_murhoa})] with comparable parameter truncations are displayed as empty markers. }
\label{Figure8}
\end{figure}

Although the present prototype study was not focused on practical applications, we did explore the effects of the unavoidable non-idealities, namely, the parameter truncations and material losses. As for the standard (magnetic) cloak, suitable truncation of the permittivities in (\ref{eq:end_eps_murho}) at the inner interface $r=R_1$ is necessary in view of their singular behavior [cf. (\ref{eq:sing_eps})]. As in Refs. \onlinecite{Kong2} and \onlinecite{Ruan}, in our parametric studies, we truncated the cloak shell at an interface $r=R_1+\Delta$, considering the thin annulus $R_1\le r<R_1+\Delta$ as part of the concealment region. Figure \ref{Figure8} shows, for an empty (vacuum) concealment region, the scattering width as a function of the truncation parameter $\Delta/R_1$, for various values of the loss-tangent ranging from zero to $10^{-2}$. For comparison, we also studied the response of a standard (magnetic) cloak \cite{Schurig2},
\begin{subequations}
\begin{eqnarray}
\varepsilon_{r}^{(ref)}(r)&=&\frac{r-R_1}{r},
\label{eq:eps1a}\\
\varepsilon_{\phi}^{(ref)}(r)&=&\frac{r}{r-R_1},
\label{eq:eps2a}\\
\mu_{z}^{(ref)}(r)&=& \left(\frac{R_2}{R_2-R_1}\right)^2 \frac{r}{r-R_1},
\label{eq:mu_a}
\end{eqnarray}
\label{eq:end_eps_murhoa}
\end{subequations}
likewise truncated so as to guarantee comparable variation ranges of the permittivities \cite{footnote2}.  
As one can observe, in spite of its non-magnetic character, the proposed cloak turns out to outperform the standard one (in terms of smaller scattering width) over a wide parametric range extending up to moderate values of the truncation parameter and losses. 
%
\begin{figure}
\begin{center}
\includegraphics[width=8cm]{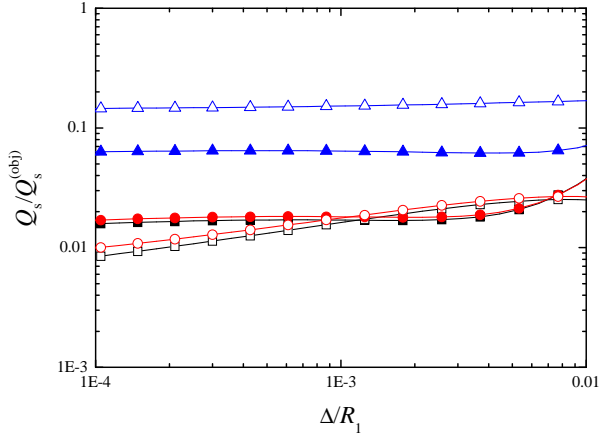}
\end{center}
\caption{(Color online) As in Fig. \ref{Figure8}, but with the concealment region entirely filled by a dielectric object with relative permittivity $\varepsilon_{obj}=4$. In order to better visualize the cloaking effect, the scattering width is normalized with respect to reference value $Q_s^{(obj)}$ exhibited by the object in vacuum.}
\label{Figure9}
\end{figure}

%
\begin{figure}
\begin{center}
\includegraphics[width=8cm]{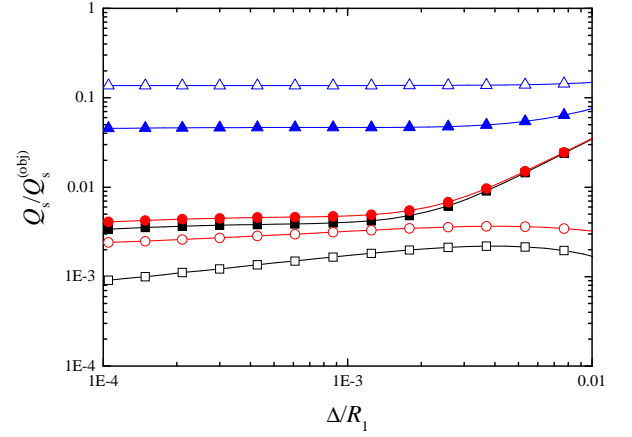}
\end{center}
\caption{(Color online) As in Fig. \ref{Figure9}, but for an object with $\varepsilon_{obj}=16$.}
\label{Figure10}
\end{figure}

%
\begin{figure}
\begin{center}
\includegraphics[width=8cm]{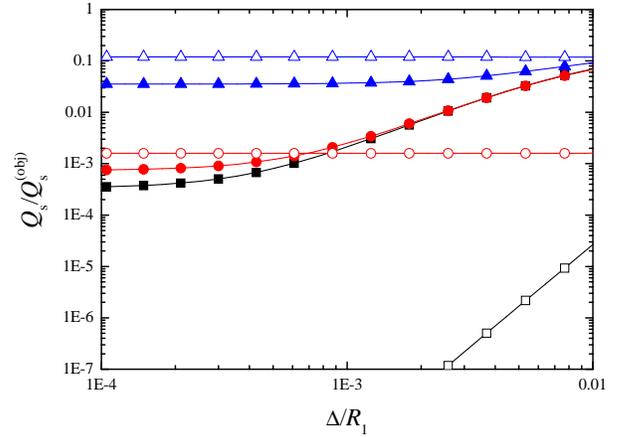}
\end{center}
\caption{(Color online) As in Fig. \ref{Figure9}, but for a PEC object.}
\label{Figure11}
\end{figure}

\begin{figure}
\begin{center}
\includegraphics[width=8cm]{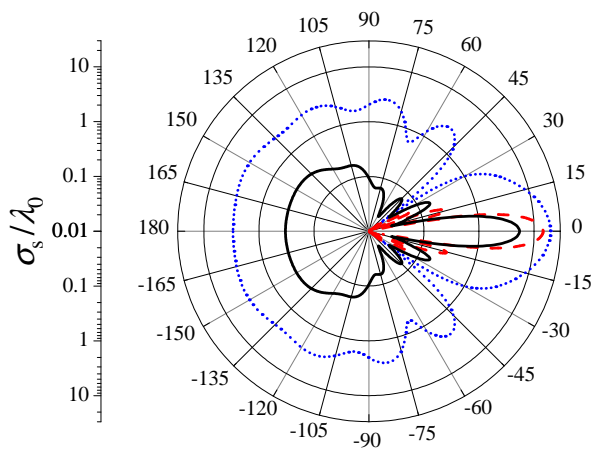}
\end{center}
\caption{(Color online) Bistatic scattering width in (\ref{eq:RCS}) (normalized with respect to the vacuum wavelength), for a concealment domain filled by a PEC object. Object cloaked via the proposed (black-solid; $Q_s/\lambda_0=0.32$) and standard (red-dashed; $Q_s/\lambda_0=0.41$) approach, respectively, with truncation parameter $\Delta/R_1=10^{-2}$ and loss-tangent $\tan\delta=10^{-2}$. Also shown, as a reference, is the response pertaining to the object free-standing in vacuum (blue-dotted; $Q_s^{(obj)}/\lambda_0=3.46$).}
\label{Figure12}
\end{figure}

It is worth pointing out that parameter truncations allow field penetration through the cloak shell, and thus, unlike in the ideal case, the overall response generally depends on the possible presence of objects in the concealment region. We therefore studied some representative scenarios, with the concealment region entirely filled up by a dielectric (with relative permittivity $\varepsilon_{obj}=4$ and $16$) or a perfect electric conductor (PEC), still amenable to analytic solution via straightforward generalization of (\ref{eq:Ht2}). Figures \ref{Figure9}--\ref{Figure11} show the corresponding responses, where, in order to better visualize the cloaking effect, the scattering width has been normalized with respect to the reference value $Q_s^{(obj)}$ exhibited by the object in vacuum. For all cases, it can be observed that the performance of proposed non-magnetic cloak is comparable to that of the standard cloak over wide parametric ranges. The sensible differences that may exist in certain parametric configurations (e.g., the PEC case in Fig. \ref{Figure11} for very small values of the truncation parameter $\Delta/R_1$ and no losses) tend to disappear when moderate values of the truncation parameter and loss-tangent are considered. In particular, for $\tan\delta=10^{-2}$, the proposed non-magnetic cloak is still capable of reducing the scattering width of over an order of magnitude with respect to the uncloaked case, slightly outperforming a comparably-truncated standard cloak. As an example, Fig. \ref{Figure12} compares the responses pertaining to a PEC cylinder free-standing in vacuum and cloaked with the proposed and standard approaches, for a truncation parameter $\Delta/R_1=10^{-2}$ and loss tangent $\tan\delta=10^{-2}$, in terms of the bistatic scattering width
\begin{eqnarray}
\sigma_s(\phi)&=&\lim_{r\rightarrow \infty}2\pi r\frac{\left|H_z^s(r,\phi)\right|^2}{\left|H_z^i(r,\phi)\right|^2}\nonumber\\
&=&\frac{4}{k_0}\left|\sum_{m=-\infty}^{\infty}(-i)^m c_m \exp\left(i m \phi\right)\right|^2,
\label{eq:RCS}
\end{eqnarray}
which describes the angular distribution of the total scattering width in (\ref{eq:Qs}). While the total scattering width values for the proposed and standard cloaks turn out to be comparable ($Q_s/\lambda_0=0.32$ and $0.41$, respectively, with a reduction of about an order of magnitude as compared to the uncloaked case), certain differences are evident in the angular distributions. In particular, as compared to the uncloaked case, our proposed cloak attains a rather uniform reduction (of almost an order of magnitude, in the worst case), whereas the 
standard cloak seems to work much better for backscattering ($\phi=180^o$) but rather poorly in the forward ($\phi=0$) direction.

%
Similar results, not shown here for brevity, were obtained for different frequencies and shape factors $R_1/R_2$.

To sum up, when the unavoidable parameter-truncation and loss effects are taken into account, the overall performance of our proposed strategy turns out to be comparable to that of a standard (magnetic) cloak. Moreover, it should be mentioned that the theoretically frequency-independent behavior of this latter is severely restricted by the unavoidable dispersion effects in any practical metamaterial implementation, so that the inherently {\em narrow-band} character of our proposed design (optimized to work at a given frequency) does not seem to constitute a critical limitation.

\section{Conclusions and Outlook}
\label{Sec:Conclusions}

In this paper, we have presented an alternative approach to non-magnetic coordinate-transformation-based invisibility cloaking. Unlike other approaches in the literature, the proposed strategy does not rely on approximate parameter reductions but rather on the design, via parametric optimization, of a {\em nearly-transparent} anisotropic and spatially inhomogeneous virtual domain, and is amenable to {\em exact analytic} treatment.

After derivation of the relevant analytic solutions, we have presented a body of representative parametric studies. Our results indicate that, when the unavoidable non-idealities (parameter truncations, losses, dispersion) of typical metamaterial implementations are taken into account, the overall performance attainable is comparable to that of a standard (magnetic) cloak.

The idea underlying the proposed strategy is rather general and, besides the cloaking scenarios, it may open up interesting perspectives for other transformation-optics applications, such as hyperlensing \cite{Kildishev1}. In this framework, it should also be emphasized that the class of constitutive relationships in (\ref{eq:start_eps_mu}) represents only one example of nearly-transparent media amenable to analytic solutions, and exploration of further classes is certainly worth of interest and currently being pursued. In particular, configurations featuring a larger number of parameters may be utilized, together with more sophisticated optimization strategies, in order to achieve broadband or multi-band responses, and/or to enforce constraints in the variation ranges of the constitutive parameters of the transformation medium.
Of particular interest are also the ``masking'' application scenarios \cite{Teixeira,Ozgun}, where one is interested in changing the scattering signature of an object (e.g., making it appear larger, smaller, or of different shape), and thus the desired scattering response is {\em inherently non-zero}. In such scenarios, the virtual-domain medium may offer extra degrees of freedom exploitable for the design of the desired scattering signature.

\appendix

\section{Pertaining to Eq. (\ref{eqSolTM1})}
\label{Sec:AppA}

Particularizing the Helmholtz equation in (\ref{eq:Helmholtz1}) to the constitutive parameters in (\ref{eq:start_eps_mu}), we obtain
\begin{eqnarray}
\!\!\!\!\!\!\!\!\!\!\!\!\!\!\!\!\!\!\!&&\left[
\frac{d^2}{d r'^2}
+\left(
\frac{1+\gamma}{r'}
+
\frac{\alpha}{R_2}
\right)
\frac{d}{d r'}\right]\Psi_m(k_0 r')
\nonumber\\
\!\!\!\!\!\!\!\!\!\!\!\!\!\!\!\!\!\!\!\!&~~~&+
\left[
k_0^2
(1-p)
+
\frac{p k_0^2 R_2}{r'}
-\left(\frac{m}{r'}\right)^2
\right]
\Psi_m(k_0 r')
=0,
\end{eqnarray}
which, letting $\tau= \xi r'$ [with $\xi$ defined in (\ref{eq:xi})], becomes
\begin{eqnarray}
\!\!\!\!\!\!\!\!\!\!\!\!\!\!\!\!\!\!\!&&\left[
\frac{d^2}{d \tau^2}
+
\left(
\frac{\alpha}{\xi R_2}
+
\frac{1+\gamma}{\tau}
\right)
\frac{d}{d \tau}\right]
\Psi_m\!\!\left(\frac{k_0 \tau}{\xi}\right)
\nonumber\\
\!\!\!\!\!\!\!\!\!\!\!\!\!\!\!\!\!\!\!\!&~~~&+\left[
\frac{(1-p)k_0^2}{\xi^2}
+
\frac{p k_0^2 R_2}{\xi \tau}-\frac{m^2}{\tau^2}
\right]
\Psi_m\!\!\left(\frac{k_0 \tau}{\xi}\right)
=0.
\label{eq:Helmholtz2}
\end{eqnarray}
and, via the mapping 
\beq
\Psi_m\!\!\left(\frac{k_0 \tau}{\xi}\right)
= \tau^{\frac{\nu_m-\gamma}{2}} 
\exp
\left[
-
\frac{(\xi R_2 +\alpha)\tau}{2 \xi R_2}
\right] \bar{\Psi}_m(\tau),
\eeq
reduces to 
\beq
\left\{
\tau
\frac{d^2}{d \tau^2}
+
\left(
1+\nu_m
-
\tau
\right)
\frac{d}{d \tau}
-
\zeta_m
\right\}
\bar{\Psi}_m(\tau)
=0,
\label{eq:Kummer}
\eeq
with $\nu_m$ and $\zeta_m$ defined in (\ref{eq:num}) and (\ref{eq:zetam}), respectively. Equation (\ref{eq:Kummer}) is readily recognized to be the Kummer equation [cf. Eq.(13.1.1) in Ref. \onlinecite{Abramowitz}], and admits two independent solutions in terms of confluent hypergeometric functions [cf. Eqs. (13.1.2), (13.1.3) in Ref. \onlinecite{Abramowitz}], from which the final solutions in (\ref{eqSolTM1}) follow straightforwardly via inverse mapping.  
Note that, as a possible alternative route, the mapping
\beq
\Psi_m\!\!\left(\frac{k_0 \tau}{\xi}\right)
= \tau^{-\left(\frac{\gamma+1}{2}\right)} 
\exp
\left(
-
\frac{\alpha\tau}{2 \xi R_2}
\right) 
\tilde{\Psi}_m(\tau)
\eeq
would lead to the Whittaker equation [cf. Eq. (13.1.31) in Ref. \onlinecite{Abramowitz}].

\section{Pertaining to the Approximations in (\ref{eq:CloakTotappenSolAA})}
\label{Sec:AppB}
We start recalling that, for $r'\rightarrow 0$, the wavefunctions $\Psi_m^{(1)}$ in (\ref{eqSolTM1}) are regular, since $\nu_m\ge\gamma$ [cf. (\ref{eq:num})] and $M(\zeta_m,\nu_m+1,0)=1$ [cf. Eq. (13.1.2) in Ref. \onlinecite{Abramowitz}], whereas $\Psi_m^{(2)}$ are singular, since [cf. Eq. (13.1.3) in Ref. \onlinecite{Abramowitz}]
\begin{eqnarray}
\!\!\!\!\!\!\!\!\!\!\!\!\!\!\!\!\!\!&&U(\zeta_m,\nu_m+1,\xi r')\sim\nonumber\\
\!\!\!\!\!\!\!\!\!\!\!\!\!\!\!\!\!\!&&~~~~~~~ -\frac{\pi (\xi r')^{-\nu_m}}{\sin[\pi (\nu_m+1)] \Gamma(\zeta_m)\Gamma(1-\nu_m)},~~r'\rightarrow 0,
\label{eq:Uapp}
\end{eqnarray}
where $\Gamma(\cdot)$ is the Gamma function [cf. Eq. (6.1.1) in Ref. \onlinecite{Abramowitz}].
We then define
\beq
\Delta'=f(R_1+\Delta),
\label{eq:Delta1}
\eeq
where, in view of (\ref{eq:g_app}),
\beq
\Delta = g(\Delta')-R_1\sim \frac{\exp(-\alpha) \chi R_2^{s-\gamma}}{R_1(\gamma-s+2)}(\Delta')^{\gamma-s+2}.
\label{eq:DD}
\eeq
Thus, combining (\ref{eqSolTM1}), (\ref{eq:Uapp}) and (\ref{eq:Delta1}), we obtain
\begin{subequations}
\begin{eqnarray}
\!\!\!\!\!\!\!\!\!\Psi_m^{(1,2)}(k_0\Delta')&\sim&
\varsigma_{1,2}
\left(
\frac{
\Delta'
}
{
R_2
}
\right)^{\frac{\pm\nu_m-\gamma}{2}},\\
\!\!\!\!\!\!\!\!\!\dot\Psi_m^{(1,2)}(k_0\Delta')
&\sim&
\varsigma_{1,2}
\left(
\frac{\pm\nu_m-\gamma}{2}\right)
\left(
\frac{
\Delta'
}
{
R_2
}
\right)^{\frac{\pm\nu_m-\gamma-2}{2}},
\end{eqnarray}
\label{eq:ppsi12}
\end{subequations}
where $\varsigma_{1,2}$ are irrelevant constants. 
Then, by recalling (\ref{eq:psi12}) and that
\beq
\dot\psi_m^{(1,2)}(k_0r)=\frac{\dot\Psi_m^{(1,2)}(k_0r')}{{\dot g}(r')},
\eeq
and substituting into (\ref{eq:CloakTot}), we obtain
\begin{subequations}
\begin{eqnarray}
d_m J_m(k_0 R_1)\!\!&\sim&\!\! a_m \varsigma_1
\left(
\frac{
\Delta'
}
{
R_2
}
\right)^{\frac{\nu_m-\gamma}{2}}\nonumber\\
\!\!&+&\!\!
b_m 
\varsigma_2
\left(
\frac{
\Delta'
}
{
R_2
}
\right)^{\frac{-\nu_m-\gamma}{2}},
\label{eq:Cloak_3aappen}
\\
d_m {\dot J_m(k_0 R_1)} 
\!\!&\sim&\!\!
\left(\displaystyle{\frac{\Delta'}{R_2}}\right)^{\gamma+1}
\left[
a_m 
\eta_1
\left(
\displaystyle{\frac{
\Delta'
}
{
R_2
}}
\right)^{\frac{\nu_m-\gamma-2}{2}}\right.\nonumber\\
\!\!&+&\!\!
\left.
b_m 
\eta_2
\left(
\frac{
\Delta'
}
{
R_2
}
\right)^{\frac{-\nu_m-\gamma-2}{2}}
\right],
\label{eq:Cloak_4aappen}
\end{eqnarray}
\label{eq:CloakTotappen}
\end{subequations}
where $\eta_{1,2}$ are other irrelevant constants.
Equations (\ref{eq:CloakTotappen}) can be readily solved with respect to $b_m$ and $d_m$, yielding
\begin{subequations}
\begin{eqnarray}
\!\!\!\!\!\!\!b_m
\!&\sim&\!
a_m\!
\left(\displaystyle{\frac{\Delta'}{R_2}}\right)^{\!\!\!\nu_m}\nonumber\\
&\times&\!\!\!\!
\left[
\frac{
\eta_1
J_m(k_0 R_1)
\!\!\left(\displaystyle{\frac{\Delta'}{R_2}}\right)^{\!\!\!\gamma}
\!\!-\!
\varsigma_1
{\dot J_m(k_0 R_1)}
}
{
\varsigma_2 {\dot J_m(k_0 R_1)}
-
\eta_2
J_m(k_0 R_1)
\left(\displaystyle{\displaystyle{\frac{\Delta'}{R_2}}}\right)^{\gamma}
}
\!\right],
\label{eq:Sol_4aappen}\\
\!\!\!\!\!\!\!d_m
\!&\sim&\!
\frac{
a_m (\eta_1 \varsigma_2-\eta_2 \varsigma_1)
\left(\displaystyle{\frac{\Delta'}{R_2}}\right)^{\frac{\gamma+\nu_m}{2}}
}
{
\varsigma_2 {\dot J_m(k_0 R_1)}
-
\eta_2
J_m(k_0 R_1)
\left(\displaystyle{\frac{\Delta'}{R_2}}\right)^{\gamma}
},
\label{eq:sol_3aappen}
\end{eqnarray}
\label{eq:CloakTotappenSol}
\end{subequations}
from which, neglecting the higher-order terms in $(\Delta'/R_2)$ and recalling (\ref{eq:DD}), the approximations in (\ref{eq:CloakTotappenSolAA}) follow straightforwardly.


\newpage


\begin{thebibliography}{999}

\bibitem{Kahn}{W. K. Kahn and H. Kurss, {\em IEEE Trans. Antennas Propag.} {\bf 13} 671, 1965.}

\bibitem{Kerker}{M. Kerker, {\em J. Opt. Soc. Am.} {\bf 65}, 376 (1975).}

\bibitem{Chew}{H. Chew and M. Kerker, {\em J. Opt. Soc. Am.} {\bf 66}, 445 (1976).}

\bibitem{Alexopoulos}{N. G. Alexopoulos and U. K. Uzunoglu, {\em Appl. Opt.} {\bf17}, 235 (1978).}

\bibitem{Kildal}{P.-S. Kildal, A. Kishk, and A. Tengs, {\em IEEE Trans. Antennas Propagat.} {\bf 44}, 1509 (1996).}

\bibitem{Hoenders}{B. J. Hoenders, {\em J. Opt. Soc. Am. A} {\bf 14}, 262 (1997).}

\bibitem{Schurig2}{D. Schurig, J. J. Mock, B. J. Justice, S. A. Cummer, J. B. Pendry, A. F. Starr, and D. R. Smith, {\em Science} {\bf 314}, 977 (2006).}

\bibitem{Alu1}{A. Al\`u and N. Engheta, {\em Phys. Rev. E} {\bf 72}, 016623 (2005).}

\bibitem{Silveirinha}{M. G. Silveirinha, A. Al\`u, and N. Engheta, {\em Phys. Rev. E} {\bf 75}, 036603 (2007).}

\bibitem{Pendry}{J. B. Pendry, D. Schurig, and D. R. Smith, {\em Science} {\bf 312}, 1780 (2006).}

\bibitem{Leonhardt1}{U. Leonhardt and T. G. Philbin, {\em New J. Phys.} {\bf 8}, 247 (2006).}

\bibitem{Leonhardt2}{U. Leonhardt, {\em Science} {\bf 312}, 1777 (2006).}

\bibitem{Schurig1}{D. Schurig, J. B. Pendry, and D. R. Smith, {\em Opt. Express} {\bf 14}, 9794 (2006).}

\bibitem{Smolyaninov}{I. I. Smolyaninov, Y. J. Hung, and C. C. Davis, {\em Opt. Lett.} {\bf 33}, 1342 (2008).}

\bibitem{Milton}{G. W. Milton and N. A. P. Nicorovici, {\em Proc. R. Soc. London A} {\bf 462}, 3027 (2006).}

\bibitem{Hakansson}{A. Hakansson, {\em Opt. Express} {\bf 15}, 4328 (2007).}

\bibitem{Alitalo}{P. Alitalo, O. Luukkonen, L. Jylha, J. Venermo, and S. A. Tretyakov, {\em IEEE Trans. Antennas Propagat.} {\bf 56}, 416 (2008).}
 
\bibitem{Alu2}{A. Al\`u and N. Engheta, {\em J. Opt. A} {\bf 10}, 093002 (2008).}

\bibitem{Leonhardt3}{U. Leonhardt and D. R. Smith, {\em New J. Phys.} {\bf 10}, 115019 (2008).}

\bibitem{Greenleaf2}{A. Greenleaf, Y. Kurylev, M. Lassas, and G. Uhlmann, {\em SIAM Rev.} {\bf 51}, 3 (2009).}

\bibitem{Greenleaf}{A. Greenleaf, Y. Kurylev, M. Lassas, and G. Uhlmann, {\em Phys. Rev. Lett.} {\bf 99}, 183901 (2007).}

\bibitem{Chen3}{H. Chen, X. Luo, H. Ma, and C. T. Chan, {\em Opt. Express} {\bf 16}, 14603 (2008).} 

\bibitem{Castaldi}{G. Castaldi, I. Gallina, V. Galdi, A. Al\`u, and N. Engheta, {\em Opt. Express} {\bf 17}, 3101 (2009).}

\bibitem{Li}{J. Li and J. B. Pendry, {\em Phys. Rev. Lett.} {\bf 101}, 203901 (2008).}

\bibitem{Liu}{R. Liu, C. Ji, J. J. Mock, J. Y. Chin, T. J. Cui, and D. R. Smith, {\em Science} {\bf 323}, 366 (2009).}

\bibitem{Gabrielli}{L. H. Gabrielli, J. Cardenas, C. B. Poitras, and M. Lipson, 	arXiv:0904.3508v1 [physics.optics] (2009).}

\bibitem{Valentine}{J. Valentine, J. Li, T. Zentgraf, G. Bartal, and X. Zhang, 	arXiv:0904.3602v1 [physics.optics] (2009).}

\bibitem{Lai}{Y. Lai, H. Chen, Z.-Q. Zhang, and C. T. Chan, {\em Phys. Rev. Lett.} {\bf 102}, 093901 (2009).}

\bibitem{Ma}{H. Ma, S. Qu, Z. Xu, and J. Wang, {\em Appl. Phys. Lett.} {\bf 94}, 103501 (2009).}

\bibitem{Cummer1}{S. A. Cummer and D. Schurig, {\em New J. Phys.} {\bf 9}, 45 (2007).}

\bibitem{Pendry2}{J. B. Pendry and J. Li, {\em New J. Phys.} {\bf 10}, 115032 (2008).}

\bibitem{Milton2}{G. W. Milton, M. Briane, and J. R. Willis, {\em New J. Phys.} {\bf 8}, 248 (2006).}

\bibitem{Zhang1}{S. Zhang, D. A. Genov, C. Sun, and X. Zhang, {\em Phys. Rev. Lett.} {\bf 100}, 123002 (2008).}

\bibitem{Cummer2}{S. A. Cummer, B.-I. Popa, D. Schurig, D. R. Smith, and J. Pendry, {\em Phys. Rev. E} {\bf 74}, 036621 (2006).}

\bibitem{Kong2}{B. Zhang, H. S. Chen, B. I. Wu, Y. Luo, L. X. Ran, and J. A. Kong, {\em Phys. Rev. B} {\bf 76}, 121101 (2007).}

\bibitem{Ruan}{Z. Ruan, M. Yan, C. W. Neff, and M. Qiu, {\em Phys. Rev. Lett.} {\bf 99}, 113903 (2007).}

\bibitem{Kong1}{H. S. Chen, B. I. Wu, B. Zhang, and J. A. Kong, {\em Phys. Rev. Lett.} {\bf 99}, 063903 (2007).}

\bibitem{Chen2}{H. Y. Chen, Z. X. Liang, P. J. Yao, X. Y. Jiang, H. R. Ma, and C. T. Chan, {\em Phys. Rev. B} {\bf 76}, 241104 (2007).}

\bibitem{Yao}{P. J. Yao, Z. X. Liang, and X. Y. Jiang, {\em Appl. Phys. Lett.} {\bf 92}, 031111 (2008).}  

\bibitem{Yan}{M. Yan, Z. C. Ruan, and M. Qiu, {\em Phys. Rev. Lett.} {\bf 99}, 233901 (2007).}

\bibitem{Tao}{H. Tao, N. I. Landy, K. Fan, A. Strikwerda, W. J. Padilla, R. D. Averitt, and X. Zhang, Proc. 2008 Int. Electron Devices Meeting (IEDM '08), San Francisco, CA, USA, Dec. 15-17, 2008, pp. 11.6.1-11.6.4.}

\bibitem{Zhou}{J. Zhou, Th. Koschny, M. Kafesaki, E. N. Economou, J. B. Pendry, and C. M. Soukoulis, {\em Phys. Rev. Lett.} {\bf 95}, 223902 (2005).}

\bibitem{Cai1}{W. S. Cai, U. K. Chettiar, A. V. Kildishev, and V. M. Shalaev, {\em Nature Photonics} {\bf 1}, 224 (2007).}

\bibitem{Cai2}{W. Cai, U. K. Chettiar, A. V. Kildishev, V. M. Shalaev, and G. W. Milton, {\em Appl. Phys. Lett.} {\bf 91}, 111105 (2007).}

\bibitem{Cai3}{W. Cai, U. K. Chettiar, A. V. Kildishev, and V. M. Shalaev, {\em Opt. Express} {\bf 16}, 5444 (2008).} 

\bibitem{Gallina}{I. Gallina, G. Castaldi, and V. Galdi, {\em Microwave Opt. Technol. Lett.} {\bf 50}, 3186 (2008).}

\bibitem{Zhang}{L. Zhang, M. Yan, and M. Qiu, {\em J. Opt. A} {\bf 10}, 5 (2008).}

\bibitem{Jacob}{Z. Jacob and E. E. Narimanov, {\em Opt. Express} {\bf 16}, 4597 (2008).} 

\bibitem{Luo}{Y. Luo, J. Zhang, H. Chen, S. Xi, and B.-I. Wu, {\em Appl. Phys. Lett.} {\bf 93}, 033504 (2008).}

\bibitem{Abramowitz}{M. Abramowitz and I. A. Stegun, {\em Handbook of Mathematical Functions} (Dover, New York, 1964).}

\bibitem{footnote1}{Obviously, the scattering width in (\ref{eq:Qs}) admits a trivial (zero-valued) global minimum for $\alpha=\gamma=p=0$ [for which the virtual-domain medium in (\ref{eq:start_eps_mu}) reduces to vacuum], which is, however, incompatible with the assumed constraints.}

\bibitem{footnote2}{Note that, in view of the different expressions, identical truncation parameters $\Delta/R_1$ in (\ref{eq:end_eps_murho}) and (\ref{eq:end_eps_murhoa}) would yield different variation ranges of the constitutive parameters. In order to guarantee a meaningful comparison, the  value of $\Delta/R_1$ reported in the graphs is referred to (\ref{eq:end_eps_murhoa}), and the one pertaining to (\ref{eq:end_eps_murho}) is adjusted accordingly so as to yield comparable (within a $\pm10\%$ window) variation ranges of the permittivities.}

\bibitem{Kildishev1}{A. V. Kildishev and E. E. Narimanov, {\em Opt. Lett.} {\bf 32}, 3432 (2007).}

\bibitem{Teixeira}{F. L. Teixeira, {\em Microwave Opt. Technol. Lett.} {\bf 49}, 2051 (2007).}

\bibitem{Ozgun}{O. Ozgun and M. Kuzuoglu, {\em Microwave Opt. Technol. Lett.} {\bf 49}, 2386 (2007).}


\end{thebibliography}
\end{document}